\documentclass[
 twocolumn,
 amsmath,amssymb,
 aps,
 pre,floatfix
]{revtex4-2}
\usepackage{url}
\usepackage[colorlinks=true, allcolors=blue]{hyperref}
\usepackage{float}
\usepackage[utf8]{inputenc}
\usepackage[T1]{fontenc}
\usepackage{lineno}
\usepackage{amsthm}
\usepackage{}

\usepackage{nicematrix}
\usepackage{bm}
\usepackage{dsfont}
\usepackage{amsfonts}
\usepackage{indentfirst}
\usepackage{graphicx}
\usepackage{dcolumn}
\usepackage{bm}
\usepackage{color}
\usepackage{wasysym}

\begin{document}

\title{Opinion dynamics of uncertain individuals}

\author{Ricardo Sim\~ao}%
 \email{ojogodascontasdevidro@gmail.com}
 

\author{Lucas Wardil}
\email{wardil@fisica.ufmg.br}
 
 \affiliation{Departamento de F\'\i sica, Universidade Federal de Minas Gerais,
Caixa Postal 702, CEP 30161-970, Belo Horizonte - MG, Brazil.
}%
\date{\today}

\begin{abstract}

The distrust in information institutions strongly affects the opinion formation process. The misinformation epidemic has the negative effect that people are unsure what to believe. In the usual opinion dynamic models, the agents have well-defined opinions that change due to social influence. Here, we present a simple model to study the emergence of consensus in opinion pools of uncertain agents that display binary opinions with a probability that depends on an individual intrinsic probability and a social term that couples the individual to the neighborhood. In regular lattices, we observe three outcomes depending on the coupling strength and the intrinsic probability: consensus is reached even in infinity systems, the agents never reach consensus, or local consensus, for both opinions, is transitory. We model individual variability by assigning different intrinsic probabilities to each agent. In this new environment, we observe regions in the space of parameters where consensus (bulk-stable, clusters of like-minded opinions) dominates. We also find weak-consensus domination regions (bulk-unstable, temporary clusters where contrary opinions emerge inside the cluster randomly) and distrust domination regions (random orientations that do not form clusters).

\end{abstract}


\maketitle

\section{Introduction}

The worldwide emergence of exacerbated opinion polarization is an important social phenomenon. The emergence of very closed-minded groups such as neo-fascists and anti-vaccination \cite{gravelle2022estimating,kaplan1998emergence} is related to polarization. Contemporary studies \cite{SM1,SM2,SM3,SM4,SM6} relate the opinion polarization to the diffusion of manipulated or incorrect information (``fake news''), and exposure algorithms in social media. On the other hand, strong distrust of government-regulated information sources is generally correlated to low-level education in \cite{ dean1961alienation}. 

Opinion dynamics has attracted the interest of physicists, who usually contribute to the area with simple models aiming at complex features that emerge from simple rules. One successful example is the voter model, whose origin is rooted in Cliﬀord and Sudbury's work on species competition in 1973 \cite{clifford1973model}, later used to study social systems in 1975 \cite{holley1975ergodic} and named ``voter model''. This model became well-known among physicists because it is one of the few non-equilibrium stochastic processes that present exact solutions in any dimension \cite{castellano2009statistical,redner2001guide}. Each agent has an associated binary variable $s = \pm 1$ representing binary opinions. At each time-step one agent $i$ and one of its neighbors $j$ are randomly selected so that agent $i$ copies the neighbor's opinion:$s_i = s_j$. This step is repeated until the system reaches the absorbing consensus ($s_i = s_j$ for all $i$ and $j$). The main question addressed by the voter model is whether consensus is reached in infinite systems.   General solutions in lattices \cite{frachebourg1996exact}, and random walkers that coalesce upon encounters theory \cite{liggett1985interacting,liggett1999stochastic} show that only at $d \le 2$ the voter model undergoes coarsening process leading to consensus. Because bulk noise is absent, disordered initial conditions lead to increasing order, indicating a relation to coarsening processes, although the roughness of the interfaces \cite{bray2002theory} suggests an unusual instance of such processes. Nevertheless, at $d>2$ the voter model  (asymptotically) presents a finite density of interfaces, implying consensus absence. This is contradictory in finite systems where the voter model reaches consensus in any dimension, and the question of interest becomes the time needed for reaching it. This time, $T_N$, is of the order of $T_N \propto O(N^2)$ for uni-dimensional systems, $T_N \propto O(N\ln N)$ for bi-dimensional systems and $T_N \propto O(N)$ for $d>2$ \cite{cox1989coalescing}.

In the voter model, the agents' opinions remain the same until they are changed due to the influence of neighbors. However, opinion uncertainty becomes especially important when distrust in information sources is present. Hence, we propose representing the agent's opinion by a random variable assuming the values $-1$ or $1$ with probabilities depending on an intrinsic term and a social term that considers the neighborhood's opinion. The agents interact with their first neighbors, and the neighbors' opinions are linearly coupled to the agents' probabilities. We are interested in the long-time behavior of the system, the conditions for the existence of the stationary state, and the microscopic mechanism driving the system. The phase diagram of the homogeneous system, where all agents have the same intrinsic probability, exhibits regions where consensus is both an attracting and absorbing state, regions where consensus is an absorbing state but not an attracting one, and finally, regions where consensus is not an attracting state nor an absorbing one. The phase diagram of heterogeneous systems has a more complex morphology. It presents the three previously discussed global states and four other regions exhibiting the coexistence of two or three of those states.

We present the model in section \textbf{II}. The results section (section \textbf{III}) is divided into three subsections discussing homogeneous systems, heterogeneous systems, and non-linear couplings. We present discussions and conclusions in the section \textbf{IV}. Additionally, the \textbf{appendix} section directs the reader to the source code (in FORTRAN 95) of the simulations presented in this work and to the calculations of particular case that can be solved using Markov chain theory.

\section{The model}

The opinion of the $i$-th agent, at time $t$, is a binary random variable $X_i^t = \pm 1$. The probability that $X_t^i=1$ is given by:
\begin{equation}\label{01}
    p_t^i = p_0^i + \frac{K}{d}\sum_{j\in \Omega_i}X_{t-1}^j, 
\end{equation}
where $\Omega_i$ is the set of the agent's first neighbors, $p_0^i$ is its intrinsic probability (baseline probability or solitude probability), $K$ is the coupling constant, and $d$ is the coordination number of the lattice. Because $p_t^i$ is a probability measure, if  $p_t^i < 0$ or $p_t^i > 1$ the value of $p_t^i$  must be truncated in the interval $[0,1]$. The update of the probabilities is synchronous. 

The model is a Markov chain in the space of states $\{-1,1\}^N$, where $N$ is the population size. Because for $N>2$ analytical solutions are unfeasible (see the appendix), we use mean-field approximations and microscopic mechanisms analysis to investigate the system. We also use computer simulations to confirm the results. The analytical solution for the case $N=2$ is presented in the appendix.

It is worth mentioning that, at $p_0^i=0.5$ and $K=0.5$, our model has the same transition rates as that of the voter model:
\begin{equation}\label{02}
\begin{split}
    W(X_t^i \rightarrow - X_{t+1}^i) 
     = & \\
     \frac{1}{2} + (\frac{1}{2}-p_0^i)X_{t-1}^i - &\frac{K}{d}X_{t-1}^i \sum_{j\in \Omega_i}X_{t-1}^j
\end{split}
\end{equation}
Notice that, in the voter model, the opinions can change only if there is no local consensus and the opinion update is asynchronous.  In our model, the opinions are intrinsically probabilistic, and the update is synchronous.

\section{Results}

\subsection{Homogeneous systems}

In homogeneous systems, all agents have the same baseline probability, $p_0^i=p_0 \; \forall i$. The definition in Eq. \ref{01} implies that $p_t^i$ takes values on a discreet spectrum ranging from $p_0-k$ to $p_0+k$ in step-sizes that depends on the coordination number $d$. For example, if $d=2$ the possible values are $\{p_0-k, p_0, p_0+k\}$. 

\subsubsection{The consensus region}
\begin{figure}
    \centering
    \includegraphics[scale = 0.6]{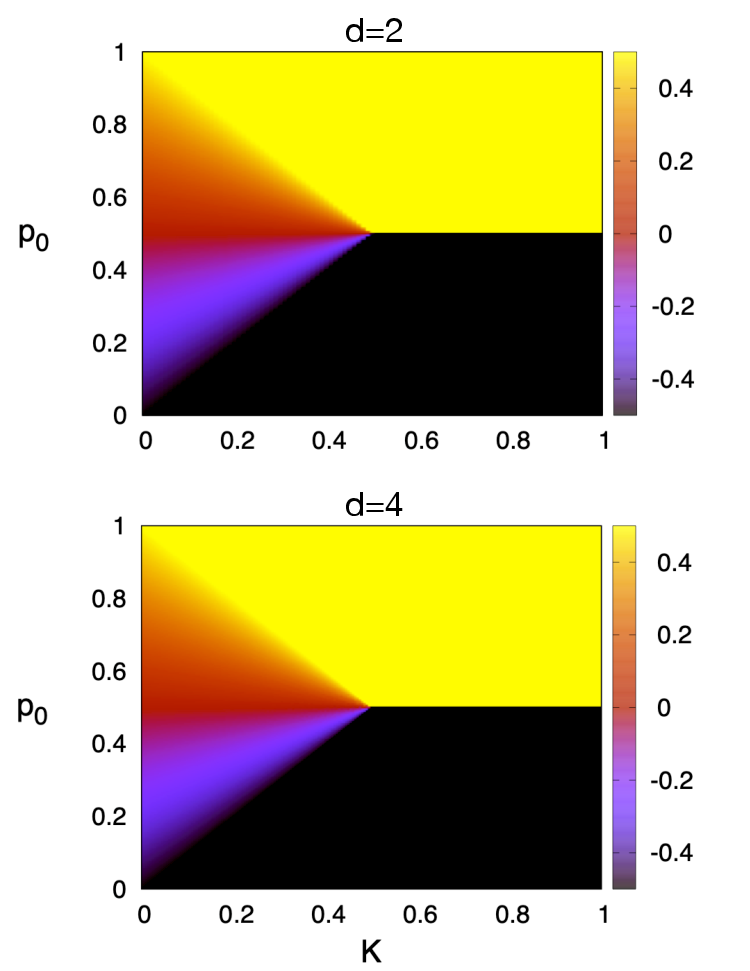}
    \caption{Phase diagrams of homogeneous systems. The heat map shows the value of $p_{\infty}-p_c$ in the stationary state averaged over all agents as a function of the intrinsic probabilities, $p_0$, and the coupling strength, $K$. The systems have $100$ agents under periodic boundary conditions and random initial opinions. The general aspect of the phase diagram does not depend on the coordination number, $d$.
    }
    \label{fig01}
\end{figure}

Figure \ref{fig01} shows the average value $p_{\infty}$ that is obtained by averaging $p_t^i$ over all agents in the lattice in the stationary state. The parameter space that we investigate is the space $p_0 \times K$.  The system reaches full consensus for $K<0.5$ in the regions above the line $p_0+K > 1$ and below the line $p_0< K$. For $K>0.5$, all points except the line $p_0 = 0.5$ yield consensus as an absorbing and attracting state.  Because the region in the parameter space where the consensus $X_t = -1$ if found is symmetric to the region $X_t = 1$, we analyse only the consensus region of $X_t = 1$. 

First, let us analyse the region ${K<0.5}$ and ${p_0+K > 1}$ and suppose a focal agent  is inside a cluster where all  display the opinion $X_t=1$ at time $t$. Inside such a cluster, because $p_0+K>1$ we must have $p_t = 1$. Thus, no fluctuation of opinions occurs. We say that this cluster is \textit{bulk-stable}. At the borders of the cluster, approximately half of the agents' neighbors are inside the bulk and half outside, in patches of agents with random orientations (if $p_0+K > 1$, then $1 - p_t = q_0-K < 0$, which implies that the probability of forming a $X_t = -1$ bulk-stable cluster is null). Therefore, at the borders $P(X_t=1)\ge p_0>0.5$ and the probability of staying at $X_t=1$ is higher than changing to $X_t=-1$. The probability that the cluster grows is higher than that it shrinks. The system reaches the consensus eventually, and $P(X_t=1)=1$ for all agents after that. For $K>0.5$ and $p_0 > 0.5$, the process of consensus formation is similar. Notice that this argument suggests that the consensus state is absorbing and attracting.

These results are also compatible with the following mean-field analysis. Let us  approximate $X_{t-1}^j$ by its average value:
\begin{eqnarray*}
E(X_{t-1}^j) &=& p_{t-1}^j - (1-p_{t-1}^j) \\
&=& 2(p_{t-1}^j-p_c),
\end{eqnarray*}
where $p_c=0.5$. This approximation converges to the exact value as $d \rightarrow \infty$. The parameter $p_c$ is a important one because it splits the parameter space in two symmetric regions. Using this average value in Eq. \ref{01} we obtain: 
\begin{equation}\label{03}
    p_t^i \approx p_0^i + \frac{2K}{d}\sum_{j\in \Omega_i}(p_{t-1}^j - p_c).
\end{equation}
If we develop this recursive expression in powers of $K$, at the limit ${t\rightarrow \infty}$ we get:
\begin{equation}\label{04}
p_{\infty}= \bigg{\{}
        \begin{array}{cc}
             p_c+\frac{p_0-p_c}{1-2K} & \textrm{ for } K<0.5  \\
             \pm \infty &  \textrm{ for } K\ge 0.5, \mbox{  and } p_0\ne p_c 
        \end{array}.
\end{equation}
We must truncate Eq. \ref{04} to have the correct probability measure. Notice that this approximation suggests the existence of the parameter $K_c = 0.5$, which is critical in the sense that the behavior of the system changes completely when the coupling $K$ crosses this value. For $K>K_c$, the opinion that reach consensus depends on whether $p_0>p_c$ or $p_0<p_c$. For $K<K_c$, the truncation of  $p_{\infty}$ yields two truncation lines: $K=p_0$, under which $p_{\infty} = 0$, and $K=1-p_0$, above which $p_{\infty} = 1$. Notice that these are the same lines obtained previously through microscopic arguments.

\begin{figure}
    \centering
    \includegraphics[scale=0.7]{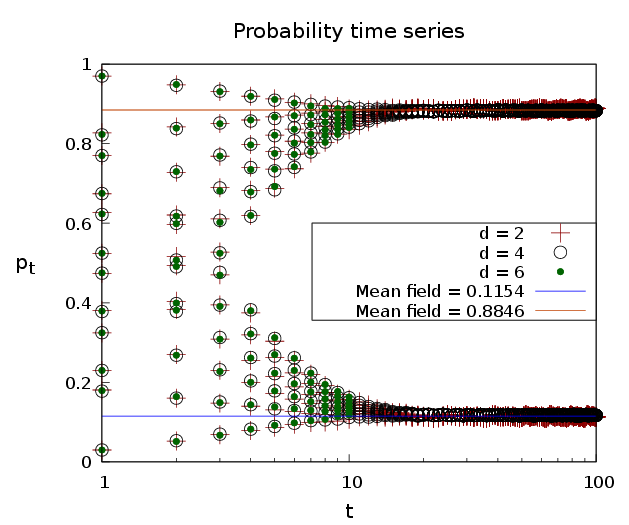}
    \caption{Time series of the average probability of the system initialized at $K=0.37$ with $p_0=0.4$ (down curves) and $p_0=0.6$ (up curves). We simulated the three geometries and also present the mean-field calculated value for comparison. We initialized the system in six different initial conditions. We initialized the opinion's initial configurations as random variables acquiring the value $X = 1$ with probability given by $p_{in}$ in the set $p_{in} = [0, 0.2, 0.4, 0.6, 0.8, 1]$. Notice that two of this configurations are consensus, and the convergence to the mean field result does not depend on these conditions. As $d$ increases, the system diverges less from the mean field.}
    \label{fig02}
\end{figure}

Figure \ref{fig02} shows time-series of the evolution of the probability (averaged over the population) in environments with $10.000$ agents in a ring ($d=2$), a torus ($d=4$) and hyper-torus ($d=6$) geometries. These results are compatible with the results obtained by transition matrix calculations (see appendix) and the mean field approximation generating Eq. \ref{04}.

\subsubsection{The consensus absence region}

In the region enclosed by the two truncation lines --  $p_0<1-K$ and $p_0>K$ -- Eq.  \ref{04} shows that $ 0 < p_{\infty} < 1$. In other words, the consensus is never reached. The agents express, with non-null probability, opinions that do not agree with the opinions of their neighbors. This happens because, independently of the opinion, any agent inside any cluster can display an opinion contrary to any local temporary consensus. In this case, no bulk-stable cluster is formed. Because the behavior is symmetric by node label exchange, any pair of nodes must have the same stationary statistical distribution, have the same opinion distributions, and be under the influence of neighboring micro-states that also have the same distribution. In other words, if the $i$-th node displays a negative opinion during an average time of $1/p_{\infty}$ before changing to the opposite opinion, then this must be true for every other node. Therefore, large clusters only appear due to strong fluctuations.

\begin{figure*}
    \centering
    \includegraphics[scale=0.35]{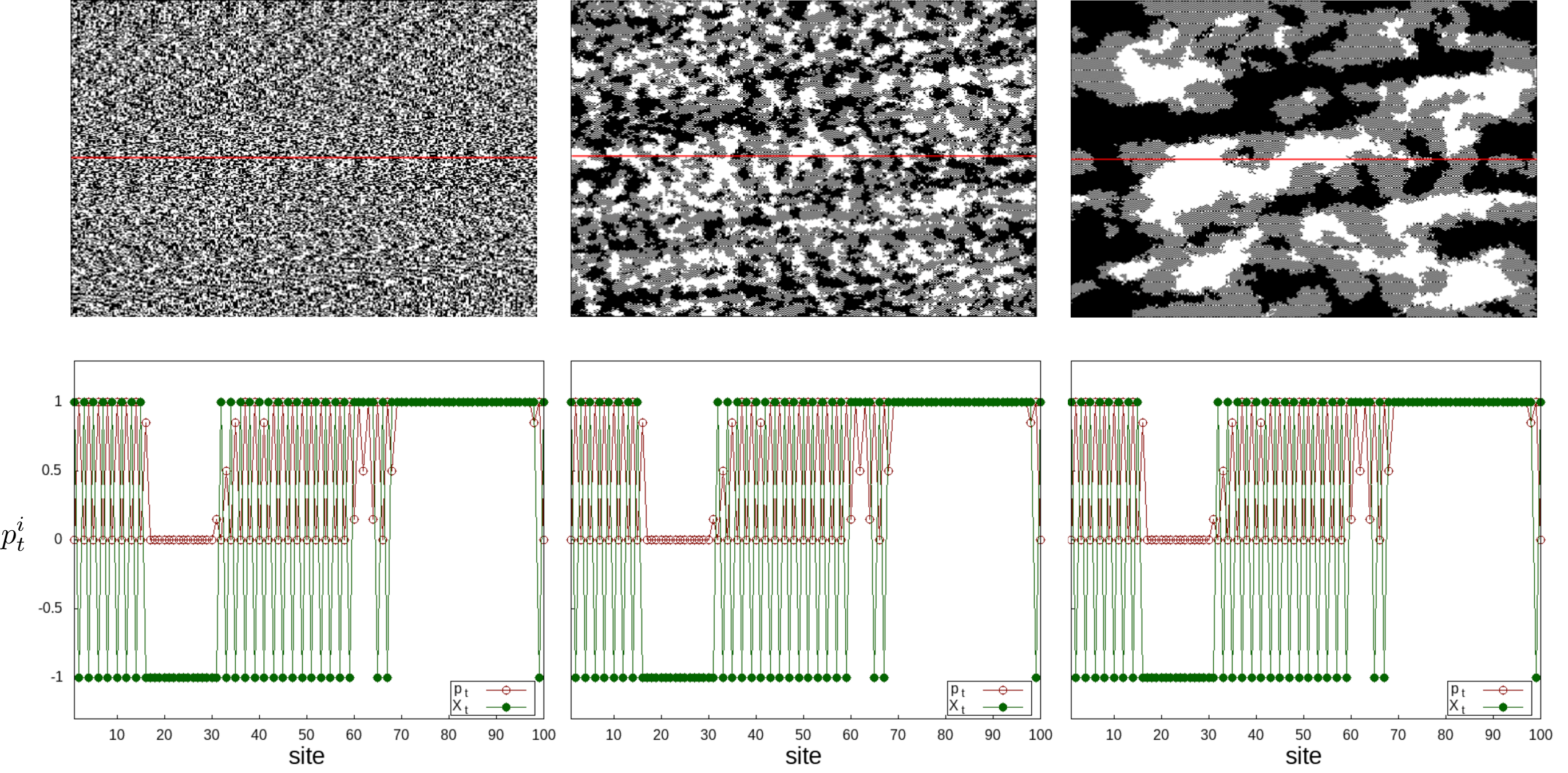}
    \caption{Time evolution of the spatial distribution of $X_t^i$ for a population on a $330 \times 300 $, $d=4$ lattice presented in three snap-shots. The upper panels show the spatial distribution of $X_t^i$ at $t=10$ (left), $t = 30$ (middle) and $t=100$ (right). Sites with $X_t^i=1$ are black and sites with $X_t^i=-1$ are white. The grey regions are chessboard patterns related to the oscillatory clusters. The lower panels show the values of $X_t^i$ and $p_t^i$ for sites at the cross-section indicated by the red line in the corresponding upper panels. Inside the stable clusters, the probability is in phase with the result, but inside the oscillatory cluster, the probability is out-of-phase with it. The parameters are $p_0=p_c$ and $K=0.7$.}
    \label{fig03}
\end{figure*}

\subsubsection{The absorbing but not attracting consensus state}

For $p_0 = p_c = 0.5$ and $K<0.5$, the probability $p_t^i$ is such that $p_t^i = p_c$ for all agents. The individuals' opinions always change, but all nodes will display both opinions with the same frequency. In this case, the conservation of orientation ($E(X_{t}^i)=0$) is just a statistical number.  

The line $p_0= p_c =0.5$, $K > K_c = 0.5$ splits the parameter space into two consensus regions. Recall that at $p_0 = p_c$ the transition rates of our model and the voter model are the same. However, the possible stationary states are very different. The consensus states $X_t=1$ and $X_t=-1$ are absorbing, but not attracting. There is another absorbing state characterized by an oscillatory pattern, which arises at the frontiers between local clusters of type $X_t=1$ and  $X_t=-1$. Interestingly, the long-time behavior of the system under random initial conditions exhibits meta-stable coexistence of all these bulk-stable states in such a way that the average probability $\left<p_t^i\right>$ is null, which is in full agreement with the mean-field approximation in Eq. \ref{04}. Nevertheless, strong local fluctuations are common, as shown in Fig. \ref{fig03}.   

\subsection{The heterogeneous system}

\begin{figure}
    \includegraphics[scale = 0.6]{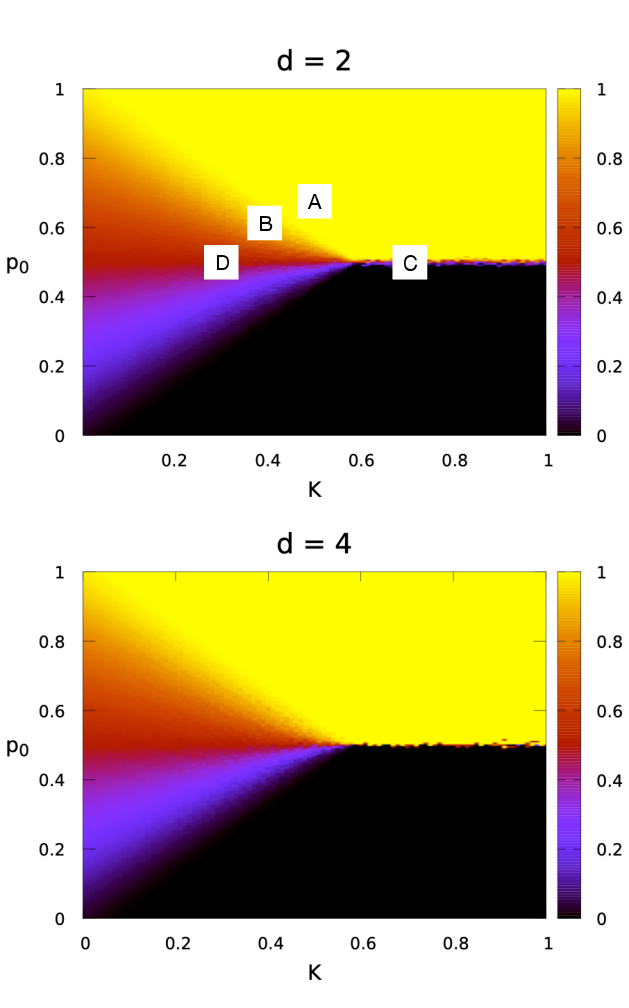}
        \caption{Fundamental diagram of the heterogeneous system in a square lattice (d=2). The parameters and interpretation are the same as in Fig. \ref{fig01}. Notice that the truncation lines are blurred, reflecting the individual variation on the value of $p_t^i$ of size $\epsilon$.  }
        \label{fig04}
\end{figure}

In heterogeneous systems, each agent has a different baseline probability: $p_0^i=p_0+\epsilon_i$. Here we consider ${\epsilon_i= \epsilon(2u_i - 1)}$, where $u_i$ is a random variable with uniform distribution in the range ${-1\le u_i\le 1}$, and $\epsilon$ measures the noise amplitude. Figure \ref{fig04} shows the phase diagram for the heterogeneous system with $\epsilon = 0.1$.  Although the phase diagram is three-dimensional, $p_0 \times K \times \epsilon$, we restrict our analyses to the two-dimensional $p_0\times K$ space, considering $\epsilon$ as a fixed parameter. The general aspect of the diagram is identical to the homogeneous case shown in Fig. \ref{fig01} except for the blurred region f linear size $\epsilon$ along the separation lines. The same microscopic arguments used to identify the consensus regions in the homogeneous case can be used here. The condition $p_0+K>1$ changes to $p_0+K+\epsilon>1$  for $K<K_c$, and the condition $p_0>0.5$ changes to $p_0+\epsilon>0.5$.

In contrast to homogeneous systems, the agents are not symmetric, and the dynamics strongly depends on the local fluctuations of $p_0^i$. Hence, a mean-field approximation is not the right approach. Instead, we will use microscopic arguments to analyze cluster stability and growth rates in the rather simple ring topology. The simulations do not indicate that different mechanisms work at higher coordination numbers.

Let us consider a three-site cluster  such that  $X^{i-1}_0 = X^{i}_0 = X^{i+1}_0 = 1$, with $X^{i+2}_0 =X^{i-2}_0 = -1$ defining the limits of the cluster ($t=0$ for simplicity). The probability that the central node and the site at the border keep their orientations is :
\begin{eqnarray}
    P(X^{i}_{1} = 1) &=& p_0 + \epsilon_i + K \;\;\mbox{ and } \label{05} \\
    P(X^{i+1}_{1} = 1) &=& p_0 + \epsilon_{i+1}. \label{06}
\end{eqnarray}
The probability that its neighbor at $i+2$ flips to the same orientation of the cluster is given by:
\begin{equation}\label{07}
    P(X^{i+2}_{1} = 1) = p_0 + \epsilon_{i+2}+K P(X^{i+3}_{0}=1).
\end{equation}
The central node keeps its orientation with probability $1$ either if 
\begin{equation}
K\ge K_c \textrm{\ and \ } p_0 + \epsilon_i > p_c, \label{cond1}
\end{equation}
or if 
\begin{equation}
K < K_c \textrm{\ and \ } p_0 + \epsilon_{i}>1-K. \label{cond2}
\end{equation}

If one of the conditions \ref{cond1} or \ref{cond2} holds for all agents, then all system have baseline probabilities lying outside of the triangular region of the parameter space and the equations \ref{06} and \ref{07} imply that $P(X^{i+1}_{1} = 1) > p_c$ and $P(X^{i+2}_{1} = 1) \ge p_0 + \epsilon_{i+2} > p_c$. Thus, the probability for the cluster to grow is higher than the shrinking probability. It will increase in size, taking over the system analogously to homogeneous systems. The consensus state is, once again, both attracting and absorbing. Notice that if $X^{i+2}_{1} = 1$ the $(i+1)$-th node becomes a bulk-node just like the $i$-th node, becoming frozen (bulk-stable).

If none of the conditions \ref{cond1} or \ref{cond2} hold for all nodes, it is still possible to have bulk-stable clusters around the nodes that satisfy them. The maximum size of such clusters depends on the baseline probabilities of the sites at the frontiers. As equations \ref{06} and \ref{07} suggests, if $p_0 + \epsilon_{i+1} + K < 1 $, than this $i$-th agent is not bulk-stable, establishing the limits of the cluster in this direction. Notice that one may observe a temporary increase in size due to fluctuations, but it is not a piece of the bulk-stable cluster as the orientation is never frozen, and contrary opinions appear in a time scale of the order of $1/(1 - (p_0+ \epsilon_{i+1}+K))$. Because the model is symmetric with respect to the transformation $p_t \leftarrow 1-p_t$ together with $X_t \leftarrow -X_t$, the same is true for the symmetrical region (a reflection of the phases diagram around the line $p_0 = p_c$). Notice that, for a fixed value of $K$, the number of not frozen sites in the positive opinion position increases as the number of nodes violating the conditions \ref{cond1} and \ref{cond2} increases.  

If no agent satisfies the conditions \ref{cond1} or \ref{cond2}, then the opinions are never frozen in clusters. All agents display frequencies of opinions compatible with their intrinsic probability. This is very similar to the inner region of the triangle in the homogeneous systems.

\begin{figure}
    \centering
     \includegraphics[scale = 0.35]{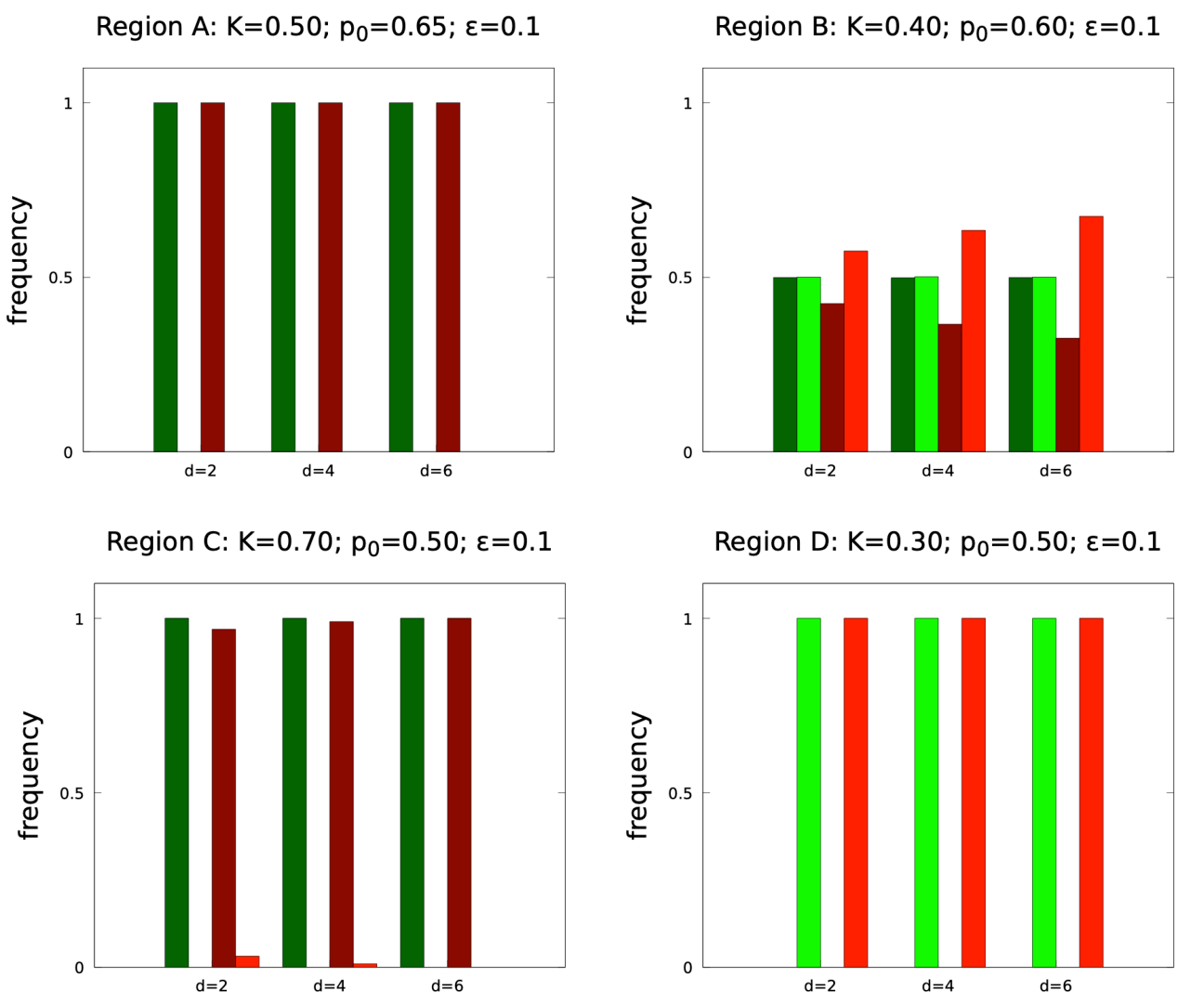}
    \caption{Histograms of the frequency of the bulk-stable and bulk-unstable clusters in regions A, B, C, and D are marked in the phases diagram. We compare the theoretical value (the number of nodes satisfying $p_0^i + K \ge 1$) with the experimental values (all nodes with $p_t^i = 1$ at the last step of the simulation). The theoretical numbers are colored green and the observed in simulations are colored red. Dark color indicate bulk-stable nodes frequency and the light colors otherwise.}
    \label{fig05}
\end{figure}

In the simulations, we counted the expected number of bulk-stable sites by counting the sites satisfying one of the conditions \ref{cond1} or \ref{cond2}. We also measured the  observed number of sites that are part of of bulk-stable clusters by counting the number of sites with $p_t = 1$ or $p_t = 0$, corresponding to $X_t = 1$ and $X_t = -1$ frozen particles, respectively, at the last simulation time. The histograms are shown  in Fig. \ref{fig05}. In the region A ($p_0 = 0.65$, $K=0.50$ and, $\epsilon = 0.1$, where all obey the conditions \ref{cond1} or \ref{cond2}), and D ($p_0 = 0.50$, $K=0.30$ and, $\epsilon = 0.1$, where none obey neither conditions \ref{cond1} or \ref{cond2}), there is an excellent agreement between the expected and the observed bulk-stable nodes. However, the agreement is not good at the mixed regions B ($p_0 = 0.60$, $K=0.40$ and, $\epsilon = 0.1$, where some obey the condition \ref{cond2}) and C ($p_0 = 0.50$, $K=0.70$ and, $\epsilon = 0.1$, where some obey the condition \ref{cond1}).

The reason for the disagreement shown in Fig. \ref{fig05} is that in the estimate of the number of agents in bulk-stable clusters, we did not include the nodes at the borders. As the coordination number increases, the fraction of frontier nodes also increases. If $K<K_c$, as in the regions B and D, to reach the bulk-stability one needs the consensus of a considerable majority, which is hindered by the increase in the coordination number (a simple majority in a low $d$ system presents a higher fluctuation than a simple majority in a high $d$ system). Therefore, the difference between the expected and the observed increases as the coordination number increases in this region. On the other hand, if $K>K_c$, as in C, the agent does not require full consensus for it to become a bulk-stable node. More nodes brings more configurations where $p_t = 1$. The probability of being in the such state becomes less dependent on fluctuations at the frontiers. Thus, the higher the coordination number, the smaller the difference.

The temporal evolution of heterogeneous systems are mostly characterized by trajectories on the basin of attraction leading to consensus states or consensus absence states. Nevertheless, complex behavior appears in the neighboring region of the separation lines in the space of parameters. The size and number distributions of stable clusters depend on the initial conditions and may increase in size beyond its limits occasionally through fluctuations, but they never dominates the system which presents frozen opinions coexisting with never-complying agents. When the parameter region contains both the truncation lines (nearby $p=pc$ and $K=K_c$ point), its temporal evolution leads to the coexistence of $X_t = 1$ bulk-stable clusters, $X_t = -1$ bulk-stable clusters, and non-conforming agents.

\subsection{Non-linear coupling}

The phase diagrams analyzed so far do not depend on the coordination number. We now show that this feature holds for general couplings taking the average over neighbors as arguments. Suppose that the coupling is of the form:

\begin{equation}\label{10}
p^i_t=p_0+Kf\left(\sum_{j\in \Omega_i}\frac{X^j_{t-1}}{d}\right),
\end{equation}
where $f(x)=x^n$ with $n$ odd, then the coordination number is not a relevant variable. To see this, let us take equation \ref{10} back in the definition \ref{01}, expand the corresponding iterated equation, approximate $X_{t-1}^j$ by its expected value, sum $p_c$ at both sides, and write $x = 2K(2(p_0-p_c))^{n-1}$. In the end, we get:

\begin{equation}\label{11}
\begin{split}
\frac{p_t-p_c}{p_0-p_c} = 1 + x(1+x(1+x(1+x...)^n)^n)^n,
\end{split}
\end{equation}
where the motive '$)^n$' repeats $t-1$ times. If $n=1$, then $x=2K$ and $(1+x(1+x(1+x...))) = \sum_{j=0}^{t}x^j$, leading us back to \ref{04}. Now let us call $Y=1+x(1+x(1+x(1+x...)^n)^n)^n $ and suppose that there exists an stationary solution for the probabilities in the sense that ${\lim_{t\rightarrow\infty} p_t = p}$. Going from $(p_t-p_c)/(p_0-p_c)$ to $(p_{t+1}-p_c)/(p_0-p_c)$ means that one must raise $Y$ to the $n$-th power, multiply it by $x$ and add $1$, but if the system is at the stationary state, $Y$ should be invariant under this specific chain of actions, or:

\begin{equation}\label{12}
Y = 1+xY^n.    
\end{equation}

The stationary state, if exists, must be the solution of \ref{12}. It does not depend on the coordination number anywhere (notice that if $n=1$ this expression gives $Y=1/(1-2K)$, and once again, we are back to \ref{04}) so that $d$ is not a relevant variable for homogeneous systems.

\subsection{Conclusions}

In the usual opinion dynamics models, individuals have opinions that can change due to the influence of others. Here we considered individuals that face uncertainty and do not fully trust their neighbors.

We show that if all agents have the same intrinsic opinion probability, social influence guarantees convergence on average to the same probability distribution. The system achieves the consensus if the coupling strength is strong enough or the agents are strongly biased towards one opinion (intrinsic probability value). However, at low coupling strength and low bias, the individuals' probabilities converge to the same, non-unitary value that guarantees that consensus will never happen, even in finite systems. 

In populations with heterogeneous intrinsic probability, the system also reaches the consensus for parameters similar to those in homogeneous systems. In heterogeneous systems, there are regions where the steady-state consists of a stable coexistence of bulk-stable clusters, temporary clusters, and agents that manifest opinions at random without belonging to a temporary cluster for a considerable time. This state is compatible with the existence of very closed-minded groups, manifesting opposite opinions, coexisting with people that do not feel confident in either opinion and people that listen to both sides, voting with the local majority. Surveys observed the coexistence of these behaviors in political contexts \cite{ternullo2022m}.

We also show that some non-linear coupling functions present results that do not depend on the coordination number $d$. At first, one may ask if such couplings would return the same results regardless of the geometry of the network (whether it is regular or not). Further studies on non-linear couplings and complex networks are required. 

Surveys in natural populations only capture the instantaneous opinion of the subjects. Hence it is hard to distinguish certain from uncertain opinions unless the degree of certainty is part of the survey. When distrust of media is widespread, feeling little confidence in an opinion is an important phenomenon to consider. Here we show that social influence and certainty are fundamental to understanding opinion dynamics with depth.

\begin{acknowledgments}
The authors thank to the Brazilian agencies CNPq and Capes.
\end{acknowledgments}

\section{Appendix}

\subsection{Supplemental  information}
 The code to run the simulations is freely available at \url{https://github.com/ojoS2/StochasticProcessCouplingProject}. 

\subsection{Markov chain analysis for a system with two agents} 

The model with $N=2$ agents is a Markov Chain  $(X_t^1,X_t^2)$ with states in the set ${\{(1,1),(1,-1),(-1,1),(-1,-1)\}}$ with transition probabilities  calculated from equation \ref{01}. For example, the probability of going from $(1,-1)$ to $(1,1)$ is equal to $p_t^1p_t^2$. The transition matrix is given by 
\begin{equation}\label{A01}
\NiceMatrixOptions{code-for-first-row=\scriptstyle,code-for-last-row=\scriptstyle}
  T=\begin{pNiceMatrix}[first-row,first-col=5]
      &(-1,-1) & (-1,1) & (1,-1) & (1,1) \\
      (-1,-1) & \alpha_q^{1}\alpha_q^{2} & \alpha_q^{1}\beta_p^{2} &\beta_p^{1}\alpha_q^{2} & \beta_p^1\beta_p^2\\
      (-1, 1) & q^1_0q^2_0 & p^1_0q^2_0 & q^1_0p^2_0& p^1_0p^2_0\\
      ( 1,-1) & q^1_0q^2_0 & q^1_0p^2_0 & p^1_0q^2_0& p^1_0p^2_0\\
      ( 1, 1) & \beta_q^{1}\beta_q^{2} & \beta_q^{1}\alpha_p^{2} &\alpha_p^{1}\beta_q^{2} & \alpha_p^1\alpha_p^2\\
  \end{pNiceMatrix},
\end{equation}
where ${\alpha_p^i = p_0^i + K}$, ${\alpha_q^i = q_0^i + K}$, ${\beta_p^i = p_0^i - K}$, ${\beta_q^i = q_0^i - K}$. The stationary state is obtained by finding the eigenvectors associated to the eigenvalue one. The particular case of the homogeneous system with parameters on the critical line $p_1^0=p_2^0=0.5$ is illustrative. The stationary state is given by
\begin{equation}
    \left(\frac{1}{4(1-2k^2)},\frac{1-4k^2}{4(1-2K^2)},\frac{1-4k^2}{4(1-2K^2)},\frac{1}{4(1-2k^2)}\right).
\end{equation}
For $K<0.5$, the pair of states $(-1,-1)$ and $(1,1)$ have equal probability, as well as the pair $(-1,1)$ and $(1,-1)$. The shift of the probability is determined by $K$: the greater the $K$ the more the system stays at $(-1,-1)$ or $(1,1)$. For $K>0.5$ we must truncate $\alpha_q^i$ and $\beta_q^i$. The stationary state is ${(1/2,0,0,1/2)}$. In this case, the system is absorbed at $(-1,-1)$ or $(1,1)$ with equal probability. For any $K$ value, the average $p_i^t$ in the stationary state is always equal to zero.

In the general homogeneous system where $p_1 = p_2 = p$, the stationary state is given by
\begin{equation}\label{A02}
\chi\left[\begin{array}{c}
     \bigg{(}\frac{(q-K)(2p^2 - (p-K)(p-q))}{(p-K)(q-2K^2-(p-q)K)}-\frac{p-q}{p-K}\bigg{)}\\
      1\\    
      1\\
      \frac{2p^2 - (p-K)(p-q)}{q-2K^2-(p-q)K}.
\end{array}\right]^t,
\end{equation}
where $\chi$ is a common factor that is determined by the normalization constraint. For example, if $q=p=1/2$ then $\chi= (1-4K^2)/(4(1-2K^2))$ and we get back the symmetry previously discussed. Figure \ref{fig01} shows the stationary value of $p_t^i$ averaged over both sites for the homogeneous case. Notice that the phase diagram for $N=2$ is nearly equal to the phase diagram for large $N$, as shown in Fig. \ref{fig01}.

\begin{figure}
    \centering
    \includegraphics[scale = 0.7]{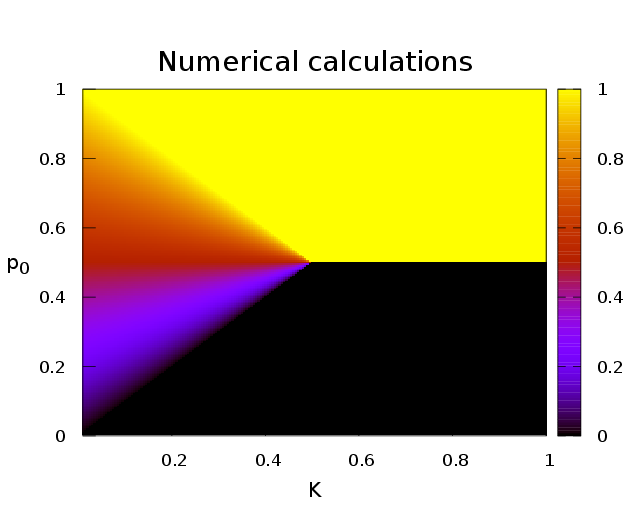}
    \caption{The phase diagram for this system calculated numerically using \ref{A02}. Compare with figure \ref{fig01}.}
    \label{figA01}
\end{figure}


%

\bibliography{references}

\begin{thebibliography}{18}%
\makeatletter
\providecommand \@ifxundefined [1]{%
 \@ifx{#1\undefined}
}%
\providecommand \@ifnum [1]{%
 \ifnum #1\expandafter \@firstoftwo
 \else \expandafter \@secondoftwo
 \fi
}%
\providecommand \@ifx [1]{%
 \ifx #1\expandafter \@firstoftwo
 \else \expandafter \@secondoftwo
 \fi
}%
\providecommand \natexlab [1]{#1}%
\providecommand \enquote  [1]{``#1''}%
\providecommand \bibnamefont  [1]{#1}%
\providecommand \bibfnamefont [1]{#1}%
\providecommand \citenamefont [1]{#1}%
\providecommand \href@noop [0]{\@secondoftwo}%
\providecommand \href [0]{\begingroup \@sanitize@url \@href}%
\providecommand \@href[1]{\@@startlink{#1}\@@href}%
\providecommand \@@href[1]{\endgroup#1\@@endlink}%
\providecommand \@sanitize@url [0]{\catcode `\\12\catcode `\$12\catcode
  `\&12\catcode `\#12\catcode `\^12\catcode `\_12\catcode `\%12\relax}%
\providecommand \@@startlink[1]{}%
\providecommand \@@endlink[0]{}%
\providecommand \url  [0]{\begingroup\@sanitize@url \@url }%
\providecommand \@url [1]{\endgroup\@href {#1}{\urlprefix }}%
\providecommand \urlprefix  [0]{URL }%
\providecommand \Eprint [0]{\href }%
\providecommand \doibase [0]{https://doi.org/}%
\providecommand \selectlanguage [0]{\@gobble}%
\providecommand \bibinfo  [0]{\@secondoftwo}%
\providecommand \bibfield  [0]{\@secondoftwo}%
\providecommand \translation [1]{[#1]}%
\providecommand \BibitemOpen [0]{}%
\providecommand \bibitemStop [0]{}%
\providecommand \bibitemNoStop [0]{.\EOS\space}%
\providecommand \EOS [0]{\spacefactor3000\relax}%
\providecommand \BibitemShut  [1]{\csname bibitem#1\endcsname}%
\let\auto@bib@innerbib\@empty
\bibitem [{\citenamefont {Gravelle}\ \emph {et~al.}(2022)\citenamefont
  {Gravelle}, \citenamefont {Phillips}, \citenamefont {Reifler},\ and\
  \citenamefont {Scotto}}]{gravelle2022estimating}%
  \BibitemOpen
  \bibfield  {author} {\bibinfo {author} {\bibfnamefont {T.~B.}\ \bibnamefont
  {Gravelle}}, \bibinfo {author} {\bibfnamefont {J.~B.}\ \bibnamefont
  {Phillips}}, \bibinfo {author} {\bibfnamefont {J.}~\bibnamefont {Reifler}},\
  and\ \bibinfo {author} {\bibfnamefont {T.~J.}\ \bibnamefont {Scotto}},\
  }\bibfield  {title} {\bibinfo {title} {Estimating the size of “anti-vax”
  and vaccine hesitant populations in the us, uk, and canada: comparative
  latent class modeling of vaccine attitudes},\ }\href@noop {} {\bibfield
  {journal} {\bibinfo  {journal} {Human vaccines \& immunotherapeutics}\
  }\textbf {\bibinfo {volume} {18}},\ \bibinfo {pages} {2008214} (\bibinfo
  {year} {2022})}\BibitemShut {NoStop}%
\bibitem [{\citenamefont {Kaplan}\ \emph {et~al.}(1998)\citenamefont {Kaplan},
  \citenamefont {Jeffrey~Kaplan}, \citenamefont {Weinberg},\ and\ \citenamefont
  {Weinberg}}]{kaplan1998emergence}%
  \BibitemOpen
  \bibfield  {author} {\bibinfo {author} {\bibfnamefont {J.}~\bibnamefont
  {Kaplan}}, \bibinfo {author} {\bibfnamefont {A.}~\bibnamefont
  {Jeffrey~Kaplan}}, \bibinfo {author} {\bibfnamefont {L.}~\bibnamefont
  {Weinberg}},\ and\ \bibinfo {author} {\bibfnamefont {L.~G.}\ \bibnamefont
  {Weinberg}},\ }\href@noop {} {\emph {\bibinfo {title} {The emergence of a
  Euro-American radical right}}}\ (\bibinfo  {publisher} {Rutgers University
  Press},\ \bibinfo {year} {1998})\BibitemShut {NoStop}%
\bibitem [{\citenamefont {Prasetya}\ and\ \citenamefont {Murata}(2020)}]{SM1}%
  \BibitemOpen
  \bibfield  {author} {\bibinfo {author} {\bibfnamefont {H.~A.}\ \bibnamefont
  {Prasetya}}\ and\ \bibinfo {author} {\bibfnamefont {T.}~\bibnamefont
  {Murata}},\ }\bibfield  {title} {\bibinfo {title} {A model of opinion and
  propagation structure polarization in social media},\ }\href@noop {}
  {\bibfield  {journal} {\bibinfo  {journal} {Computational Social Networks}\
  }\textbf {\bibinfo {volume} {7}},\ \bibinfo {pages} {1} (\bibinfo {year}
  {2020})}\BibitemShut {NoStop}%
\bibitem [{\citenamefont {Lee}\ \emph {et~al.}(2014)\citenamefont {Lee},
  \citenamefont {Choi}, \citenamefont {Kim},\ and\ \citenamefont {Kim}}]{SM2}%
  \BibitemOpen
  \bibfield  {author} {\bibinfo {author} {\bibfnamefont {J.~K.}\ \bibnamefont
  {Lee}}, \bibinfo {author} {\bibfnamefont {J.}~\bibnamefont {Choi}}, \bibinfo
  {author} {\bibfnamefont {C.}~\bibnamefont {Kim}},\ and\ \bibinfo {author}
  {\bibfnamefont {Y.}~\bibnamefont {Kim}},\ }\bibfield  {title} {\bibinfo
  {title} {Social media, network heterogeneity, and opinion polarization},\
  }\href@noop {} {\bibfield  {journal} {\bibinfo  {journal} {Journal of
  communication}\ }\textbf {\bibinfo {volume} {64}},\ \bibinfo {pages} {702}
  (\bibinfo {year} {2014})}\BibitemShut {NoStop}%
\bibitem [{\citenamefont {Bail}\ \emph {et~al.}(2018)\citenamefont {Bail},
  \citenamefont {Argyle}, \citenamefont {Brown}, \citenamefont {Bumpus},
  \citenamefont {Chen}, \citenamefont {Hunzaker}, \citenamefont {Lee},
  \citenamefont {Mann}, \citenamefont {Merhout},\ and\ \citenamefont
  {Volfovsky}}]{SM3}%
  \BibitemOpen
  \bibfield  {author} {\bibinfo {author} {\bibfnamefont {C.~A.}\ \bibnamefont
  {Bail}}, \bibinfo {author} {\bibfnamefont {L.~P.}\ \bibnamefont {Argyle}},
  \bibinfo {author} {\bibfnamefont {T.~W.}\ \bibnamefont {Brown}}, \bibinfo
  {author} {\bibfnamefont {J.~P.}\ \bibnamefont {Bumpus}}, \bibinfo {author}
  {\bibfnamefont {H.}~\bibnamefont {Chen}}, \bibinfo {author} {\bibfnamefont
  {M.~F.}\ \bibnamefont {Hunzaker}}, \bibinfo {author} {\bibfnamefont
  {J.}~\bibnamefont {Lee}}, \bibinfo {author} {\bibfnamefont {M.}~\bibnamefont
  {Mann}}, \bibinfo {author} {\bibfnamefont {F.}~\bibnamefont {Merhout}},\ and\
  \bibinfo {author} {\bibfnamefont {A.}~\bibnamefont {Volfovsky}},\ }\bibfield
  {title} {\bibinfo {title} {Exposure to opposing views on social media can
  increase political polarization},\ }\href@noop {} {\bibfield  {journal}
  {\bibinfo  {journal} {Proceedings of the National Academy of Sciences}\
  }\textbf {\bibinfo {volume} {115}},\ \bibinfo {pages} {9216} (\bibinfo {year}
  {2018})}\BibitemShut {NoStop}%
\bibitem [{\citenamefont {Tucker}\ \emph {et~al.}(2018)\citenamefont {Tucker},
  \citenamefont {Guess}, \citenamefont {Barber{\'a}}, \citenamefont {Vaccari},
  \citenamefont {Siegel}, \citenamefont {Sanovich}, \citenamefont {Stukal},\
  and\ \citenamefont {Nyhan}}]{SM4}%
  \BibitemOpen
  \bibfield  {author} {\bibinfo {author} {\bibfnamefont {J.~A.}\ \bibnamefont
  {Tucker}}, \bibinfo {author} {\bibfnamefont {A.}~\bibnamefont {Guess}},
  \bibinfo {author} {\bibfnamefont {P.}~\bibnamefont {Barber{\'a}}}, \bibinfo
  {author} {\bibfnamefont {C.}~\bibnamefont {Vaccari}}, \bibinfo {author}
  {\bibfnamefont {A.}~\bibnamefont {Siegel}}, \bibinfo {author} {\bibfnamefont
  {S.}~\bibnamefont {Sanovich}}, \bibinfo {author} {\bibfnamefont
  {D.}~\bibnamefont {Stukal}},\ and\ \bibinfo {author} {\bibfnamefont
  {B.}~\bibnamefont {Nyhan}},\ }\bibfield  {title} {\bibinfo {title} {Social
  media, political polarization, and political disinformation: A review of the
  scientific literature},\ }\href@noop {} {\bibfield  {journal} {\bibinfo
  {journal} {Political polarization, and political disinformation: a review of
  the scientific literature (March 19, 2018)}\ } (\bibinfo {year}
  {2018})}\BibitemShut {NoStop}%
\bibitem [{\citenamefont {Guerra}\ \emph {et~al.}(2013)\citenamefont {Guerra},
  \citenamefont {Meira~Jr}, \citenamefont {Cardie},\ and\ \citenamefont
  {Kleinberg}}]{SM6}%
  \BibitemOpen
  \bibfield  {author} {\bibinfo {author} {\bibfnamefont {P.}~\bibnamefont
  {Guerra}}, \bibinfo {author} {\bibfnamefont {W.}~\bibnamefont {Meira~Jr}},
  \bibinfo {author} {\bibfnamefont {C.}~\bibnamefont {Cardie}},\ and\ \bibinfo
  {author} {\bibfnamefont {R.}~\bibnamefont {Kleinberg}},\ }\bibfield  {title}
  {\bibinfo {title} {A measure of polarization on social media networks based
  on community boundaries},\ }in\ \href@noop {} {\emph {\bibinfo {booktitle}
  {Proceedings of the international AAAI conference on web and social
  media}}},\ Vol.~\bibinfo {volume} {7}\ (\bibinfo {year} {2013})\ pp.\
  \bibinfo {pages} {215--224}\BibitemShut {NoStop}%
\bibitem [{\citenamefont {Dean}(1961)}]{dean1961alienation}%
  \BibitemOpen
  \bibfield  {author} {\bibinfo {author} {\bibfnamefont {D.~G.}\ \bibnamefont
  {Dean}},\ }\bibfield  {title} {\bibinfo {title} {Alienation: Its meaning and
  measurement},\ }\href@noop {} {\bibfield  {journal} {\bibinfo  {journal}
  {American sociological review}\ ,\ \bibinfo {pages} {753}} (\bibinfo {year}
  {1961})}\BibitemShut {NoStop}%
\bibitem [{\citenamefont {Clifford}\ and\ \citenamefont
  {Sudbury}(1973)}]{clifford1973model}%
  \BibitemOpen
  \bibfield  {author} {\bibinfo {author} {\bibfnamefont {P.}~\bibnamefont
  {Clifford}}\ and\ \bibinfo {author} {\bibfnamefont {A.}~\bibnamefont
  {Sudbury}},\ }\bibfield  {title} {\bibinfo {title} {A model for spatial
  conflict},\ }\href@noop {} {\bibfield  {journal} {\bibinfo  {journal}
  {Biometrika}\ }\textbf {\bibinfo {volume} {60}},\ \bibinfo {pages} {581}
  (\bibinfo {year} {1973})}\BibitemShut {NoStop}%
\bibitem [{\citenamefont {Holley}\ and\ \citenamefont
  {Liggett}(1975)}]{holley1975ergodic}%
  \BibitemOpen
  \bibfield  {author} {\bibinfo {author} {\bibfnamefont {R.~A.}\ \bibnamefont
  {Holley}}\ and\ \bibinfo {author} {\bibfnamefont {T.~M.}\ \bibnamefont
  {Liggett}},\ }\bibfield  {title} {\bibinfo {title} {Ergodic theorems for
  weakly interacting infinite systems and the voter model},\ }\href@noop {}
  {\bibfield  {journal} {\bibinfo  {journal} {The annals of probability}\ ,\
  \bibinfo {pages} {643}} (\bibinfo {year} {1975})}\BibitemShut {NoStop}%
\bibitem [{\citenamefont {Castellano}\ \emph {et~al.}(2009)\citenamefont
  {Castellano}, \citenamefont {Fortunato},\ and\ \citenamefont
  {Loreto}}]{castellano2009statistical}%
  \BibitemOpen
  \bibfield  {author} {\bibinfo {author} {\bibfnamefont {C.}~\bibnamefont
  {Castellano}}, \bibinfo {author} {\bibfnamefont {S.}~\bibnamefont
  {Fortunato}},\ and\ \bibinfo {author} {\bibfnamefont {V.}~\bibnamefont
  {Loreto}},\ }\bibfield  {title} {\bibinfo {title} {Statistical physics of
  social dynamics},\ }\href@noop {} {\bibfield  {journal} {\bibinfo  {journal}
  {Reviews of modern physics}\ }\textbf {\bibinfo {volume} {81}},\ \bibinfo
  {pages} {591} (\bibinfo {year} {2009})}\BibitemShut {NoStop}%
\bibitem [{\citenamefont {Redner}(2001)}]{redner2001guide}%
  \BibitemOpen
  \bibfield  {author} {\bibinfo {author} {\bibfnamefont {S.}~\bibnamefont
  {Redner}},\ }\href@noop {} {\emph {\bibinfo {title} {A guide to first-passage
  processes}}}\ (\bibinfo  {publisher} {Cambridge university press},\ \bibinfo
  {year} {2001})\BibitemShut {NoStop}%
\bibitem [{\citenamefont {Frachebourg}\ and\ \citenamefont
  {Krapivsky}(1996)}]{frachebourg1996exact}%
  \BibitemOpen
  \bibfield  {author} {\bibinfo {author} {\bibfnamefont {L.}~\bibnamefont
  {Frachebourg}}\ and\ \bibinfo {author} {\bibfnamefont {P.~L.}\ \bibnamefont
  {Krapivsky}},\ }\bibfield  {title} {\bibinfo {title} {Exact results for
  kinetics of catalytic reactions},\ }\href@noop {} {\bibfield  {journal}
  {\bibinfo  {journal} {Physical Review E}\ }\textbf {\bibinfo {volume} {53}},\
  \bibinfo {pages} {R3009} (\bibinfo {year} {1996})}\BibitemShut {NoStop}%
\bibitem [{\citenamefont {Liggett}\ and\ \citenamefont
  {Liggett}(1985)}]{liggett1985interacting}%
  \BibitemOpen
  \bibfield  {author} {\bibinfo {author} {\bibfnamefont {T.~M.}\ \bibnamefont
  {Liggett}}\ and\ \bibinfo {author} {\bibfnamefont {T.~M.}\ \bibnamefont
  {Liggett}},\ }\href@noop {} {\emph {\bibinfo {title} {Interacting particle
  systems}}},\ Vol.~\bibinfo {volume} {2}\ (\bibinfo  {publisher} {Springer},\
  \bibinfo {year} {1985})\BibitemShut {NoStop}%
\bibitem [{\citenamefont {Liggett}\ \emph {et~al.}(1999)\citenamefont {Liggett}
  \emph {et~al.}}]{liggett1999stochastic}%
  \BibitemOpen
  \bibfield  {author} {\bibinfo {author} {\bibfnamefont {T.~M.}\ \bibnamefont
  {Liggett}} \emph {et~al.},\ }\href@noop {} {\emph {\bibinfo {title}
  {Stochastic interacting systems: contact, voter and exclusion processes}}},\
  Vol.\ \bibinfo {volume} {324}\ (\bibinfo  {publisher} {springer science \&
  Business Media},\ \bibinfo {year} {1999})\BibitemShut {NoStop}%
\bibitem [{\citenamefont {Bray}(2002)}]{bray2002theory}%
  \BibitemOpen
  \bibfield  {author} {\bibinfo {author} {\bibfnamefont {A.~J.}\ \bibnamefont
  {Bray}},\ }\bibfield  {title} {\bibinfo {title} {Theory of phase-ordering
  kinetics},\ }\href@noop {} {\bibfield  {journal} {\bibinfo  {journal}
  {Advances in Physics}\ }\textbf {\bibinfo {volume} {51}},\ \bibinfo {pages}
  {481} (\bibinfo {year} {2002})}\BibitemShut {NoStop}%
\bibitem [{\citenamefont {Cox}(1989)}]{cox1989coalescing}%
  \BibitemOpen
  \bibfield  {author} {\bibinfo {author} {\bibfnamefont {J.~T.}\ \bibnamefont
  {Cox}},\ }\bibfield  {title} {\bibinfo {title} {Coalescing random walks and
  voter model consensus times on the torus in zd},\ }\href@noop {} {\bibfield
  {journal} {\bibinfo  {journal} {The Annals of Probability}\ ,\ \bibinfo
  {pages} {1333}} (\bibinfo {year} {1989})}\BibitemShut {NoStop}%
\bibitem [{\citenamefont {Ternullo}(2022)}]{ternullo2022m}%
  \BibitemOpen
  \bibfield  {author} {\bibinfo {author} {\bibfnamefont {S.}~\bibnamefont
  {Ternullo}},\ }\bibfield  {title} {\bibinfo {title} {“i’m not sure what
  to believe”: Media distrust and opinion formation during the covid-19
  pandemic},\ }\href@noop {} {\bibfield  {journal} {\bibinfo  {journal}
  {American Political Science Review}\ ,\ \bibinfo {pages} {1}} (\bibinfo
  {year} {2022})}\BibitemShut {NoStop}%
\end{thebibliography}%
\end{document}